\documentclass[11pt]{article}
\usepackage{times}
\usepackage{geometry}
\usepackage[utf8]{inputenc}
\usepackage{enumitem,amssymb}
\usepackage{ragged2e}
\usepackage{graphicx}
\usepackage{comment}
\usepackage{multicol}
\usepackage{titlesec}
\titlespacing*{\section}{0pt}{0.1\baselineskip}{0.0\baselineskip}
\titlespacing*{\subsection}{0pt}{0.1\baselineskip}{0.0\baselineskip}
\usepackage[usenames]{xcolor} 
\definecolor{xlinkcolor}{cmyk}{1,1,0,0}
\usepackage{url}
\usepackage[
 colorlinks=true,    
 linkcolor=xlinkcolor,     
 citecolor=xlinkcolor,     
 filecolor=xlinkcolor,  
 urlcolor=xlinkcolor,      
 final=true
]{hyperref}
\usepackage[super,sort&compress,comma]{natbib}
\usepackage{enumitem}
\usepackage{lineno}
\setenumerate{itemsep=0mm}

\begin{document}
\begin{raggedright} 
\huge
Snowmass 2021 - Letter of Interest \hfill \\[+1em]
\textit{The Probe Of Multi-Messenger Astrophysics (POEMMA)} \hfill \\[+1em]
\end{raggedright}

\normalsize

\noindent {\large \bf Thematic Areas:}  (check all that apply $\square$/$\blacksquare$)

\noindent $\square$ (CF1) Dark Matter: Particle Like \\
\noindent $\square$ (CF2) Dark Matter: Wavelike  \\ 
\noindent $\square$ (CF3) Dark Matter: Cosmic Probes  \\
\noindent $\square$ (CF4) Dark Energy and Cosmic Acceleration: The Modern Universe \\
\noindent $\square$ (CF5) Dark Energy and Cosmic Acceleration: Cosmic Dawn and Before \\
\noindent $\blacksquare$ (CF6) Dark Energy and Cosmic Acceleration: Complementarity of Probes and New Facilities \\
\noindent $\blacksquare$ (CF7) Cosmic Probes of Fundamental Physics \\

\noindent {\large \bf Contact Information:}\\
Angela Olinto (University of Chicago) [{\tt aolinto@uchicago.edu}]\\
Fred Sarazin (Colorado School of Mines) [{\tt fsarazin@mines.edu}]\\
Collaboration: POEMMA\\
 
\noindent {\large \bf Authors:} {\raggedright A.V.Olinto, F.Sarazin, J.H.Adams, R.Aloisio, L.A.Anchordoqui, M.Bagheri, D.Barghini, M.Battisti, D.R.Bergman, M.E.Bertaina, P.F.Bertone, F.Bisconti, M.Bustamante, M.Casolino, M.J.Christl, A.L.Cummings, I.De Mitri, R.Diesing, R.Engel, J.Eser, K.Fang, G.Fillipatos, F.Fenu, E.Gazda, C.Gu\'epin, E.A.Hays, E.G.Judd, P.Klimov, J.Krizmanic, V.Kungel, E.Kuznetsov, S.Mackovjak, L.Marcelli, J.McEnery, K.-D.Merenda, S.S.Meyer, J.W.Mitchell, H.Miyamoto, J.M.Nachtman, A.Neronov, F.Oikonomou, Y. Onel, A.N.Otte, E.Parizot, T.Paul, J.S.Perkins, P.Picozza, L.W.Piotrowski, G.Prévôt, P.Reardon, M.H.Reno, M.Ricci, O.Romero Matamala, K.Shinozaki, J.F.Soriano, F.Stecker, Y.Takizawa, R.Ulrich, M.Unger, T.M.Venters, L.Wiencke, D.Winn, R.M.Young, M.Zotov \\[+1em]}


\noindent {\large \bf Abstract:}  The Probe Of Extreme Multi-Messenger Astrophysics (POEMMA)  is designed to identify the sources of Ultra-High-Energy Cosmic Rays (UHECRs) and to observe cosmic neutrinos, both with full-sky coverage. Developed as a NASA Astrophysics Probe-class mission, POEMMA consists of two spacecraft flying in a loose formation at 525 km altitude, 28.5$^\circ$ inclination orbits. Each spacecraft hosts a Schmidt telescope with a large collecting area and wide field of view. A novel focal plane is optimized to observe both the UV fluorescence signal from extensive air showers (EASs) and the beamed optical Cherenkov signals from EASs. In POEMMA-stereo fluorescence mode, POEMMA will measure the spectrum, composition, and full-sky distribution of the UHECRs above 20 EeV with high statistics along with remarkable sensitivity to UHE neutrinos. The spacecraft are designed to quickly re-orient to a POEMMA-limb mode to observe neutrino emission from  Target-of-Opportunity (ToO)  transient astrophysical sources viewed just below the Earth’s limb.  In this mode, POEMMA will have unique sensitivity to cosmic $\nu_{\tau}$ events above 20 PeV  by measuring the upward-moving EASs induced by the decay of the emerging $\tau$ leptons following the interactions of $\nu_{\tau}$ inside the Earth.


\newpage

\noindent POEMMA (Probe Of Extreme Multi-Messenger Astrophysics) is a NASA probe mission designed to observe Ultra-High-Energy Cosmic Rays (UHECRs) and cosmic neutrinos from space~\cite{Olinto:2019euf}. With its twin telescopes, POEMMA will monitor colossal volumes of the Earth's atmosphere to detect extensive air showers (EASs) produced by extremely energetic cosmic messengers: UHECRs above 20 EeV (1 EeV $\equiv$ 10$^{18}$eV) and cosmic neutrinos above 20 PeV (1 PeV $\equiv$ 10$^{15}$eV) with sensitivity over the entire sky. 

\noindent A POEMMA white paper was submitted to the Astro2020 decadal survey~\cite{Olinto:2019mjh}. Given the strong expertise of DOE-led high-energy physics projects, an appropriate DOE contribution to a joint  NASA/DOE mission would be the POEMMA hybrid focal surface detectors and electronics.

\section{The Observatory} 
The design of the POEMMA observatory 
evolved from previous work on the OWL~\cite{Stecker:2004wt} and JEM-EUSO~\cite{Adams:2013vea} designs, the CHANT concept~\cite{Neronov:2016zou}, and the sub-orbital payloads EUSO-SPB1~\cite{Wiencke:2017cfi} and EUSO-SPB2~\cite{Adams:2017fjh}. POEMMA is composed of two identical space telescopes that provide significant advantages in terms of exposure and sky coverage to sources of the highest energy particles. Each telescope consists of a wide (45$^{\circ}$) field-of-view (FoV) Schmidt optics system with a 4-meter mirror. The focal surface has two complementary capabilities: a fast (1 $\mu$s) ultraviolet camera to observe EAS fluorescence signals and an ultrafast (10 ns) optical camera to detect EAS Cherenkov signals. This hybrid camera is designed to optimize the different science objectives. EASs from UHECRs and cosmic neutrinos are observed from an orbit altitude of 525 km and a wide range of directions in the dark sky.

\noindent The scientific objectives of POEMMA are achieved by operating the telescopes in two different orientation modes: a quasi-nadir stereo fluorescence configuration, for precise UHECR observations 
(denoted POEMMA-stereo),
and a tilted, Earth-limb viewing configuration 
(denoted POEMMA-limb). 
In POEMMA-stereo mode, POEMMA is also sensitive to Ultra-High-Energy neutrinos. The POEMMA-limb mode is used to point both telescopes in the direction of astrophysical transient events rising or setting just below the Earth's limb to detect neutrino emission from astrophysical targets-of-opportunity (ToOs). It also allows for a much greater exposure to UHECRs, albeit with a higher energy threshold. 

\noindent In the POEMMA-stereo configuration, the two wide-angle telescopes, each with several square meters of effective photon collecting area, view a common, immense atmospheric volume corresponding to approximately $10^{4}$ gigatons of atmosphere. The POEMMA-stereo mode yields one order of magnitude increase in yearly UHECR exposure compared to that obtainable by ground observatory arrays and two orders of magnitude compared to ground fluorescence observations. In POEMMA-limb mode, POEMMA searches for optical Cherenkov signals of upward-moving EASs generated by tau lepton decays produced by $\nu_\tau$ interactions in the
Earth. The terrestrial neutrino target monitored by POEMMA  reaches nearly $10^{10}$ gigatons. 
In this configuration, an even more extensive volume of the atmosphere is monitored for UHECR fluorescence measurements. Thus, POEMMA uses the Earth and its atmosphere as a gargantuan high-energy physics detector and astrophysics observatory.

\section{The Scientific reach}
The main scientific goals of  POEMMA  are to discover the elusive sources of UHECRs with energies above 20 EeV and to observe cosmic neutrinos from multi-messenger transients. POEMMA exploits the tremendous gains in both UHECR and cosmic neutrino exposures offered by space-based measurements, including full-sky coverage of the celestial sphere. 

\subsection{Ultra-High-Energy Cosmic-Rays (UHECRs)}
The nature of the astrophysical sources of UHECRs and their acceleration mechanism(s) remains a mystery~\cite{Kotera:2011cp,Anchordoqui:2018qom,AlvesBatista:2019tlv,Sarazin:2019fjz}. POEMMA is designed to obtain definitive measurements of the UHECR spectrum, composition, and source locations for $E \gtrsim 20~{\rm  EeV}$, and fulfills the requirements expected from a next-generation instrument by the UHECR community~\cite{Sarazin:2019fjz}.
In both the POEMMA-stereo and POEMMA-limb configurations, EAS fluorescence signals are observed as video recordings with 1~$\mu$s snapshots.  Each POEMMA telescope records an EAS trace in its focal surface, which defines an observer-EAS plane. In POEMMA-stereo mode, the intersection of the two observer-EAS planes accurately defines the geometry of the EAS trajectory. Precise reconstruction of the EAS is achieved for opening angles between these two planes larger than $\sim5^\circ$. In POEMMA-limb observations, the EAS trajectory reconstruction is based on monocular reconstruction where the distance to the EAS in the observer-EAS plane is determined by the evolution in time of the EAS and a model of the atmosphere.

\noindent Over its planned 5-year operation, POEMMA will collect a dataset larger than the current statistics of the Pierre Auger Observatory (Auger) and the Telescope Array (TA) experiment combined~\cite{Anchordoqui:2019omw}.
With full-sky coverage, POEMMA will observe the UHECR source distribution over the full celestial sphere, eliminating the need for cross-calibration between two different experiments with only partial-sky coverage.
Together with primary composition and spectrum measurements well beyond the flux suppression \cite{Greisen:1966jv, Zatsepin:1966jv, Allard:2008gj}, POEMMA will be capable of detecting anisotropy at the level of $5\sigma$~\cite{Anchordoqui:2019omw} for cross-correlation search parameters within the vicinity of the signal regions for the anisotropy hints reported by TA~\cite{Abbasi:2014lda,Abbasi:2020fxl} and Auger~\cite{Aab:2018chp}. 
Thus, POEMMA will turn the TA and Auger anisotropy hints (and/or other anisotropy signals yet to be discovered) into significant detection to finally discover the locations of the UHECR sources.

\subsection{Cosmic neutrinos}
POEMMA will also be sensitive to the most energetic cosmic neutrinos, from 20~PeV to the ZeV scale \cite{Beresinsky:1969qj}, thus providing an opportunity to make substantial progress in high-energy astrophysics and fundamental physics \cite{SM2021LoICosmicNeutrino}. In the POEMMA-limb configurations, EAS Cherenkov signals from the decay of tau leptons as they exit the Earth's surface \cite{Fargion:2000iz,Feng:2001ue} are observed as video recordings with 10~ns snapshots. Observable tau lepton decay events for POEMMA are viewed in the directions starting close to the limb of the Earth located at 67.5$^{\circ}$ from the nadir for POEMMA's 525 km altitude.

\noindent POEMMA will be especially suited for rapid follow-up of ToOs \cite{SM2021LoIToO} via neutrinos with energies $E_{\nu} \gtrsim$ 20 PeV, because it will orbit the Earth in a period of $95$~mins and will be capable of re-pointing its satellites by $90^{\circ}$ in $500$~s in its transient tracking mode. In combination, these design features will enable POEMMA to access nearly the entire dark sky within the time scale of one orbit. 
POEMMA will also have groundbreaking sensitivity to neutrinos at energies beyond $100$~PeV, reaching the level of modeled neutrino fluences for nearby sources in many astrophysical scenarios \cite{Venters:2019xwi}. Note that POEMMA also has sensitivity to neutrinos with energies above 20~EeV through fluorescence observations of neutrino induced EASs~\cite{Anchordoqui:2019omw}. 

\subsection{Fundamental physics \& other science objectives}
{\bf pp cross section beyond collider energies:} The EASs developing in the atmosphere observed by POEMMA present a fixed-target calorimeter experiment
with $E_0 \gtrsim 20$ EeV. This provides a measure of the inelastic nuclear interactions at an effective center-of-mass energy of $\sqrt{2 E_0 m_p} \approx 280$ TeV, well above the current and future \cite{Abada2019} capabilities of the LHC.\\
\noindent {\bf Searches for superheavy dark matter (SHDM):} When SHDM decays into standard model particles, the final state products are dominated by photons and neutrinos, whose EASs are detectable by POEMMA~\cite{Alcantara:2019sco}. For neutrinos, both the Cherenkov and fluorescence signals are available yielding outstanding senstivity for $E_\nu \gtrsim 20$ PeV to above a ZeV. POEMMA's photon sensitivity for EAS fluorescence measurements provides an order of magnitude improvement \cite{Anchordoqui:2019omw} over the current limits from ground UHECR experiments.\\
\noindent {\bf Supplementary science capabilities} of POEMMA include other probes of physics beyond the Standard Model of particle physics (e.g. violation of Lorentz Invariance~\cite{Stecker:2017gdy}), the study of atmospheric transient luminous events (TLEs), and the search for meteors and nuclearites~\cite{Jorge}.

\clearpage

\bibliographystyle{utphys}
\bibliography{main.bib}

\end{document}